\documentclass[aip, jcp, reprint, onecolumn, 12pt]{revtex4-2}

\usepackage[utf8]{inputenc}
\usepackage[T1]{fontenc}
\usepackage{amsmath, amssymb, amsthm}
\usepackage{graphicx}
\usepackage{xcolor}
\usepackage{hyperref}

\begin{document}

\title{Symplectic Widths and Squeezed-State Suppression in Quantum Reaction Dynamics:\\
Covariance Geometry, Quantum Normal Forms, and Transmission Diagnostics}

\author{Stephen Wiggins}
\affiliation{Hetao Institute of Mathematics and Interdisciplinary 
Sciences, Shenzhen, China}
\affiliation{School of Mathematics, University of Bristol, UK}

\date{\today}

\begin{abstract}
Recent work in classical reaction dynamics has suggested that transport
through an index-1 saddle may be organized not only by flux, but also by
candidate local symplectic width scales associated with bounded,
full-dimensional neighborhoods of the transition-state bottleneck.  In
this paper we examine a quantum companion to this classical
symplectic-filter viewpoint.  The aim is not to prove a quantum version of
Gromov's non-squeezing theorem for reaction dynamics.  Rather, we ask how
the covariance geometry of highly squeezed Gaussian wavepackets affects
transmission across a quantum normal-form bottleneck.

The distinction between classical symplectic capacity and quantum
squeezed-state geometry is essential.  A squeezed Gaussian state is represented by a Wigner distribution on the
full phase space, with a covariance matrix that is highly anisotropic in a
chosen bath coordinate--momentum plane.  The symplectic eigenvalue of the
covariance matrix remains fixed for a pure squeezed Gaussian, whereas the
oscillator action-area diagnostic used below grows with the squeeze
parameter.  We therefore introduce a covariance-based bath-action scale,
derived from $2\pi\langle \hat J_2\rangle_s$, as a quantum diagnostic to be
compared with the classical candidate bottleneck width.  In this interpretation, transverse squeezing changes how the incoming state
occupies the bath-mode phase-space directions, increasing the expected bath
action and thereby reducing the effective energy available to the reactive
coordinate at fixed total expectation energy.

To avoid the instability of direct propagation for states with extreme
phase-space eccentricity, we work in the Weyl-symbol formulation of the
quantum normal form (QNF).  For the quadratic saddle--center model, we use
an exact separable baseline transmission formula obtained by convolving
the squeezed-state number distribution in the bath mode with the
one-dimensional Kemble transmission factor for the reactive coordinate.
For anharmonic truncated QNF models, we enforce algebraic conservation of
the total expectation energy and evaluate Gaussian expectation-value
diagnostics of the Weyl symbol using Wick--Isserlis moment formulas.

The calculations show a pronounced squeeze-induced suppression in the
transmission diagnostics.  As the squeezed-state bath-action scale grows relative to
the classical candidate width, the expected bath action increases rapidly,
the effective reactive energy is depleted, and the Kemble transmission
diagnostic enters a strongly suppressed regime.  These results support a
quantum geometric suppression mechanism consistent with the classical
candidate-width picture, while leaving open the formulation of a genuine
quantum non-squeezing theorem for reaction dynamics.  The resulting
framework links squeezed-state covariance geometry, quantum normal forms,
normal-form action scales, and finite-time transmission diagnostics near
an index-1 saddle.
\end{abstract}

\maketitle

\section{Introduction}

The phase-space formulation of transition state theory has revealed that
reaction dynamics near an index-1 saddle is organized by a hierarchy of
geometric structures: dividing surfaces with no-local-recrossing
properties, normally hyperbolic invariant manifolds (NHIMs), stable and
unstable manifolds, and directional flux through the bottleneck region.
In the classical setting these structures can be computed locally using
Poincar\'e--Birkhoff normal form theory, which provides adapted canonical
coordinates separating the reaction direction from the bounded transverse
bath modes \cite{Wiggins1994,Uzer2002,Waalkens2008}.

The accompanying classical symplectic-filter framework adds another
geometric ingredient.  Gromov's non-squeezing theorem shows that
Hamiltonian flow is constrained not only by Liouville volume preservation,
but also by a two-dimensional symplectic rigidity in canonical
coordinate--momentum planes \cite{Gromov1985}.  In reaction dynamics one
cannot assign a finite symplectic capacity directly to the fixed-energy
surface, since that surface is odd-dimensional and unbounded in the
reaction direction.  Instead, the classical theory considers bounded,
full-dimensional proxy domains or narrow energy layers near the
transition-state bottleneck, and uses normal-form bath-action geometry to
define candidate local symplectic width scales \cite{Wiggins2026_Classical}.
These candidate scales are not substitutes for flux.  They ask a different
question: how is an incoming phase-space set arranged relative to the
canonical transverse bath directions identified by the normal form?

The purpose of the present paper is to investigate a quantum counterpart
of this question.  We ask whether highly squeezed incoming quantum states
can show strong transmission suppression when their bath-mode covariance
geometry is compared with the classical candidate width scale of the
bottleneck. This comparison should therefore be read as a phase-space diagnostic for
squeeze-induced transmission suppression; it does not establish a new
non-squeezing theorem for quantum states or quantum reaction dynamics.

Several distinctions are important at the outset.  First, the classical
non-squeezing theorem is a statement about symplectic embeddings of
full-dimensional classical domains.  A Wigner distribution is a quantum
phase-space representation of a state, not a classical domain transported
by a Hamiltonian symplectomorphism.  Therefore the calculations below
should not be read as a proof of a quantum non-squeezing theorem for
chemical reactions.

Second, the quantity that grows with the squeeze parameter is not a
symplectic capacity of the quantum state.  A pure squeezed Gaussian is a
minimum-uncertainty state obtained from the unsqueezed minimum-uncertainty Gaussian state by a linear
symplectic transformation.  Its covariance ellipsoid becomes highly
anisotropic in the fixed bath coordinates, but the symplectic eigenvalue
associated with the covariance matrix remains fixed at the
minimum-uncertainty scale.  What grows is the expectation of the bath
action in the oscillator coordinates selected by the normal form.  The
quantity
\[
        2\pi\langle \hat J_2\rangle_s
\]
is therefore used below as a covariance-based bath-action area diagnostic,
not as a new definition of quantum capacity.

Third, the word ``squeezing'' has two different meanings in this paper.
Classically, Gromov non-squeezing is a theorem about the impossibility of
symplectically embedding a ball into a smaller canonical cylinder.
Quantum mechanically, a squeezed state is a Gaussian state whose covariance
matrix is anisotropic in a coordinate--momentum plane \cite{Gerry2004}.
The relation between these notions is interpretive rather than literal:
quantum squeezed states provide controlled probes of the bath-plane action
scales suggested by the classical symplectic-filter picture.

To make this comparison tractable, we use the Weyl-symbol formulation of
the quantum normal form (QNF).  Direct propagation of highly squeezed
wavepackets can be numerically delicate because one coordinate direction
is sharply localized while the conjugate direction is broadly distributed.
Grid-based propagation must resolve both scales simultaneously
\cite{Kosloff1988,Tannor2007}, while semiclassical initial value
representations can become unstable when the broad conjugate distribution
samples strongly anharmonic regions near the saddle \cite{Miller2001,Kay2005}.
The QNF formulation avoids these difficulties by replacing direct
wavepacket propagation with algebraic evaluation of Weyl-symbol
observables in adapted normal-form coordinates \cite{Waalkens2008,Lee1995}.

In the quadratic saddle--center model, the QNF is separable and yields an
exact baseline transmission formula: the squeezed-state occupation
probabilities in the bath mode are convolved with the one-dimensional
Kemble transmission factor for the reactive coordinate.  In the
anharmonic truncated-QNF setting, a closed-form multidimensional
transmission coefficient is not available.  Instead, we compute exact
Gaussian expectation-value diagnostics of the truncated Weyl symbol using
Wick--Isserlis moment formulas.  The resulting energy-depletion formula is
not itself a scattering theorem, but it is a precise diagnostic of how
transverse squeezing redistributes the expectation energy between bath and
reaction degrees of freedom.

The paper is organized as follows.  Section~II connects the classical
normal-form bottleneck with the Weyl-symbol and Moyal formulations of
quantum phase-space dynamics.  Section~III defines the squeezed-state
covariance diagnostics and separates covariance-based action scales from
symplectic capacity.  Section~IV develops the quadratic transmission
baseline and the anharmonic QNF expectation-value diagnostic.  Section~V
introduces the relative squeeze suppression metric.  Section~VI discusses
the physical interpretation, and Section~VII summarizes the conclusions
and open problems.

\section{From Classical Bottlenecks to Weyl-Symbol Quantum Dynamics}

To analyze how quantum squeezing affects transmission, we first anchor the
terminology in the classical normal-form geometry of the bottleneck.  In a
neighborhood of an index-1 saddle, classical normal-form coordinates
separate one hyperbolic reaction degree of freedom from the transverse bath
degrees of freedom.  The reactive integral $I$ is associated with motion in
the saddle direction.  The bath actions $J_k$ are associated with the
bounded oscillator degrees of freedom.  On the NHIM, the reaction variables
vanish, so the restricted dynamics is generated entirely by these bath
degrees of freedom.  The NHIM is the invariant equator of the dividing
surface, and its bath-action boundary determines the candidate transverse
width scales used in the classical symplectic-filter framework.

The corresponding quantum description is most naturally expressed in
phase space through the Weyl calculus.  The Wigner function $W(q,p,t)$ is
the phase-space representative of the quantum state, and the Weyl symbol
$H_W(q,p)$ is the phase-space representative of the Hamiltonian operator.
The exact phase-space evolution is governed by the Moyal equation
\begin{equation}
    \frac{\partial W}{\partial t}
    =\{H_W,W\}_M
    =\frac{2}{\hbar}H_W(q,p)
      \sin\left(\frac{\hbar}{2}\Lambda\right)W(q,p),
    \label{eq:Moyal}
\end{equation}
where
\begin{equation}
    \Lambda
    =\overleftarrow{\partial}_q\overrightarrow{\partial}_p
    -\overleftarrow{\partial}_p\overrightarrow{\partial}_q
\end{equation}
is the symplectic differential operator.  The arrows indicate whether the
derivative acts on the Hamiltonian symbol to the left or on the Wigner
function to the right.

Although Eq.~\eqref{eq:Moyal} is exact, it is not the most convenient
computational object for highly squeezed states.  The derivatives in the
Moyal expansion can become large in the squeezed direction and in the
extended conjugate direction.  For the local saddle problem, the QNF
provides a more stable representation.  In a neighborhood of the saddle,
the Weyl symbol of the $N$th-order QNF Hamiltonian has the form
\begin{equation}
    \mathcal{K}_{\mathrm{QNF}}^{(N)}(I,J_2,\ldots,J_n,\hbar)
    =E_0+\lambda I+\sum_{k=2}^n \omega_k J_k
     +\sum_{m=2}^N \mathcal{P}_m(I,\mathbf J,\hbar),
    \label{eq:QNF_symbol}
\end{equation}
where $\mathbf J=(J_2,\ldots,J_n)$ and the polynomials
$\mathcal P_m$ contain the higher-order quantum normal-form corrections.
The important point is that the local quantum dynamics is encoded in a
polynomial in action variables and powers of $\hbar$.  This makes it
possible to evaluate expectation-value diagnostics for Gaussian Wigner
states algebraically, without time-propagating a highly eccentric Wigner
function.

\section{Squeezed-State Covariance Geometry and Action Diagnostics}

The classical symplectic-filter paper introduces a candidate bottleneck
width scale, denoted $c_{\mathrm{cand}}(E)$, by using the maximal bath
actions permitted by the NHIM at energy $E$ \cite{Wiggins2026_Classical}.
For a two-degree-of-freedom quadratic saddle--center model, this scale is
\begin{equation}
    c_{\mathrm{cand}}(E)=2\pi J_2^{\max}(E).
    \label{eq:ccand_two_dof}
\end{equation}
For higher-dimensional normal forms, the corresponding candidate scale is
\begin{equation}
    c_{\mathrm{cand}}(E)=2\pi\min_{k\ge2}J_k^{\max}(E).
\end{equation}
These quantities are classical bath-plane action-area scales associated
with bounded, full-dimensional neighborhoods of the bottleneck.  They are
not assigned to the odd-dimensional fixed-energy surface itself.

The quantum calculation uses this classical scale as a reference.  The
incoming state is a Gaussian wavepacket aligned with the normal-form
coordinates, with one transverse bath mode prepared in a squeezed vacuum
state $|0,s\rangle$.  In the bath plane $(q_2,p_2)$, the covariance matrix
of this state is
\begin{equation}
    \Sigma_s
    =\begin{pmatrix}
        \langle \hat q_2^2\rangle_s & 0\\
        0 & \langle \hat p_2^2\rangle_s
      \end{pmatrix}
    =\frac{\hbar}{2}
      \begin{pmatrix}
        e^{-2s} & 0\\
        0 & e^{2s}
      \end{pmatrix}.
    \label{eq:covariance}
\end{equation}
The associated Wigner function is the zero-mean Gaussian
\begin{equation}
    W_s(q_2,p_2)
    =\frac{1}{\pi\hbar}
      \exp\left[-\frac{1}{\hbar}
      \left(e^{2s}q_2^2+e^{-2s}p_2^2\right)\right].
    \label{eq:wigner_squeezed}
\end{equation}
Thus increasing $s$ narrows the distribution in $q_2$ and broadens it in
$p_2$, preserving the minimum-uncertainty product.

The oscillator bath action is
\begin{equation}
        J_2=\frac{1}{2}(q_2^2+p_2^2).
\end{equation}
For the squeezed vacuum,
\begin{equation}
    \langle \hat J_2\rangle_s
    =\frac{\hbar}{2}\cosh(2s).
    \label{eq:J_exp}
\end{equation}
This is the central quantum diagnostic.  It measures how much oscillator
action the squeezed state carries in the fixed bath coordinates selected
by the normal form.

It is essential to distinguish Eq.~\eqref{eq:J_exp} from a symplectic
capacity statement.  The covariance matrix \eqref{eq:covariance} has
symplectic eigenvalue $\hbar/2$ for all $s$; the squeezed state is obtained
from the vacuum by a linear symplectic transformation.  Thus the minimum
symplectic covariance scale is fixed.  What changes with $s$ is the
expectation of the particular oscillator action $J_2$ associated with the
bath Hamiltonian.  We therefore define the covariance-based action-area
scale
\begin{equation}
    a_{\mathrm{sq}}(s)
    :=2\pi\langle \hat J_2\rangle_s
    =\pi\hbar\cosh(2s).
    \label{eq:a_sq}
\end{equation}
The notation $a_{\mathrm{sq}}(s)$ is used deliberately: this is a
bath-plane action-area diagnostic, not a symplectic capacity of the
quantum state.

\subsection{Quantum thickening and the role of uncertainty}

In the classical theory, a fixed-energy surface must be thickened to a
full-dimensional energy layer or localized proxy domain before a
non-squeezing comparison is meaningful.  The quantum situation is
different but related.  A physical quantum state cannot be localized on a
singular classical energy surface with zero width in all phase-space
directions.  Its Wigner distribution has a finite covariance structure
constrained by the uncertainty principle.  In this limited sense, the
state supplies its own phase-space thickening.

This observation should not be overinterpreted.  The Wigner distribution
of a quantum state is not a classical symplectic domain, and its covariance
ellipse is not the same object as the classical energy layer.  The point is
that the covariance matrix provides a precise full-dimensional phase-space
geometry for the incoming quantum state.  The diagnostic question is then
whether the bath-action scale associated with that covariance geometry is
large compared with the classical candidate width of the bottleneck.

\subsection{Geometric and reactive-energy depletion thresholds}

We distinguish two thresholds.  The first is a geometric diagnostic
threshold, defined by
\begin{equation}
        a_{\mathrm{sq}}(s_{\mathrm{geom}})
        =c_{\mathrm{cand}}(E).
        \label{eq:sgeom_def}
\end{equation}
This threshold says that the bath-action area scale of the squeezed state
has become comparable to the classical candidate bottleneck width.

The second is a reactive-energy depletion threshold.  For the linearized
two-degree-of-freedom saddle--center model, the effective expectation energy available
to the reaction coordinate is
\begin{equation}
    \langle \hat H_{\mathrm{react}}\rangle_s
    =E-\omega_2\langle \hat J_2\rangle_s
    =E-\frac{\hbar\omega_2}{2}\cosh(2s).
    \label{eq:Hreact_quad}
\end{equation}
The reactive-energy depletion threshold occurs when
\begin{equation}
    \langle \hat H_{\mathrm{react}}\rangle_s=0,
    \qquad
    s_{\mathrm{dep}}
    =\frac{1}{2}\cosh^{-1}\left(\frac{2E}{\hbar\omega_2}\right).
    \label{eq:sdep}
\end{equation}
The thresholds \eqref{eq:sgeom_def} and \eqref{eq:sdep} are related,
but they are conceptually distinct.  The first compares bath-action area
scales; the second records depletion of the expectation energy available
to the reactive degree of freedom.  Neither threshold is asserted to be a
rigorous quantum non-squeezing threshold.

\section{Transmission Diagnostics and Model Systems}

We now connect the covariance diagnostics to transmission.  Two levels of
analysis are used.  The first is the quadratic saddle--center model, where
the QNF is exactly separable and the squeezed-state transmission can be
computed as an exact sum over bath-mode occupations.  The second is an
anharmonic truncated-QNF model, where a full multidimensional transmission
coefficient is not computed.  In that case we use exact Gaussian
expectation values of the truncated Weyl symbol to diagnose reactive-energy
depletion.

\subsection{Quadratic separable baseline}

For the two-degree-of-freedom quadratic saddle--center model,
\begin{equation}
        \hat H=\lambda\hat I+\omega_2\hat J_2,
\end{equation}
the reactive and bath degrees of freedom are separable.  If the bath mode
occupies the eigenstate $|n\rangle$, with energy
\begin{equation}
        E_n=\hbar\omega_2\left(n+\frac12\right),
\end{equation}
then the reactive coordinate sees the reduced energy $E-E_n$.  The
one-dimensional inverted-oscillator transmission factor is the Kemble
factor \cite{Kemble1935,Barton1986}
\begin{equation}
        T_{\mathrm K}(E-E_n)
        =\left[1+\exp\left(-\frac{2\pi}{\hbar\lambda}(E-E_n)\right)
        \right]^{-1}.
\end{equation}

For a squeezed vacuum in the bath mode, the number distribution is
supported on even quantum numbers.  Writing $n=2m$, one has
\begin{equation}
    P_{2m}(s)
    =\frac{(\tanh s)^{2m}}{\cosh s}
      \frac{(2m)!}{2^{2m}(m!)^2},
    \label{eq:squeezed_number_distribution}
\end{equation}
with $P_{2m+1}(s)=0$ \cite{Gerry2004}.  The exact quadratic transmission
baseline is therefore
\begin{equation}
    T_{\mathrm{quad}}(E,s)
    =\sum_{m=0}^{\infty}P_{2m}(s)
     \left[1+\exp\left(-\frac{2\pi}{\hbar\lambda}
     \left(E-E_{2m}\right)\right)\right]^{-1}.
    \label{eq:T_quad}
\end{equation}
This formula is exact for the separable quadratic model.  It shows how
squeezing modifies transmission by redistributing probability over bath
energy levels, thereby changing the statistical energy available to the
reactive coordinate.

\subsection{Anharmonic QNF expectation-value diagnostics}

For an anharmonic truncated QNF, the exact multidimensional scattering
problem is generally not separable.  What remains exactly computable is
the expectation of polynomial Weyl symbols against Gaussian Wigner states.
This gives a controlled diagnostic of how squeezing changes the energy
budget in the normal-form approximation.

Consider the two-degree-of-freedom Weyl symbol
\begin{equation}
    H_W=E_0+\lambda I+\omega_2J_2+\alpha J_2^2+b_2 I J_2+\cdots.
    \label{eq:HW}
\end{equation}
In the bath plane,
\begin{equation}
        J_2=\frac12(q_2^2+p_2^2),
\end{equation}
so
\begin{equation}
        J_2^2=\frac14(q_2^4+2q_2^2p_2^2+p_2^4).
\end{equation}
The squeezed Wigner function is Gaussian, so its polynomial moments are
computed by Wick--Isserlis formulas \cite{Isserlis1918,Papoulis2002}.  For
independent squeezed variables,
\begin{equation}
    \langle q_2^{2m}p_2^{2\ell}\rangle_s
    =(2m-1)!!(2\ell-1)!!
      \left(\frac{\hbar}{2}\right)^{m+\ell}e^{2s(\ell-m)}.
    \label{eq:wick_moments}
\end{equation}
In particular,
\begin{equation}
    \langle \alpha J_2^2\rangle_s
    =\frac{\alpha\hbar^2}{16}
      \left(3e^{-4s}+2+3e^{4s}\right).
    \label{eq:J2_squared_expectation}
\end{equation}

We now impose the total expectation-energy constraint
\begin{equation}
        \langle H_W\rangle_s=E.
\end{equation}
For a separable incoming state, the expectation of the product term factors
as
\begin{equation}
        \langle I J_2\rangle_s=\langle I\rangle_s\langle J_2\rangle_s.
\end{equation}
This factorization is exact for the incoming product state used in the
calculation; it should not be interpreted as an exact statement about the
fully evolved interacting scattering state.  Defining

\begin{equation}
        \langle H_{\mathrm{react}}\rangle_s=\lambda\langle I\rangle_s,
\end{equation}
the energy constraint gives
\begin{equation}
    E=E_0+\langle H_{\mathrm{react}}\rangle_s
      +\omega_2\langle J_2\rangle_s
      +\alpha\langle J_2^2\rangle_s
      +\frac{b_2}{\lambda}
       \langle H_{\mathrm{react}}\rangle_s\langle J_2\rangle_s.
\end{equation}
Solving for the effective reactive energy yields
\begin{equation}
    \langle H_{\mathrm{react}}\rangle_s
    =\frac{\lambda\left(E-E_0-
      \omega_2\langle J_2\rangle_s-
      \alpha\langle J_2^2\rangle_s\right)}
      {\lambda+b_2\langle J_2\rangle_s}.
    \label{eq:H_react_exact}
\end{equation}
Equation~\eqref{eq:H_react_exact} is the principal anharmonic diagnostic.
It is exact for the truncated polynomial Weyl symbol and the chosen
Gaussian product state.  It is not an exact multidimensional transmission
coefficient.  Its role is to quantify how squeezing increases bath-action
expectations and thereby reduces the expectation energy available to the
reaction coordinate.

\begin{figure}[ht]
    \centering
    \IfFileExists{quantum_blockade_fig.png}{%
    \includegraphics[width=0.7\textwidth]{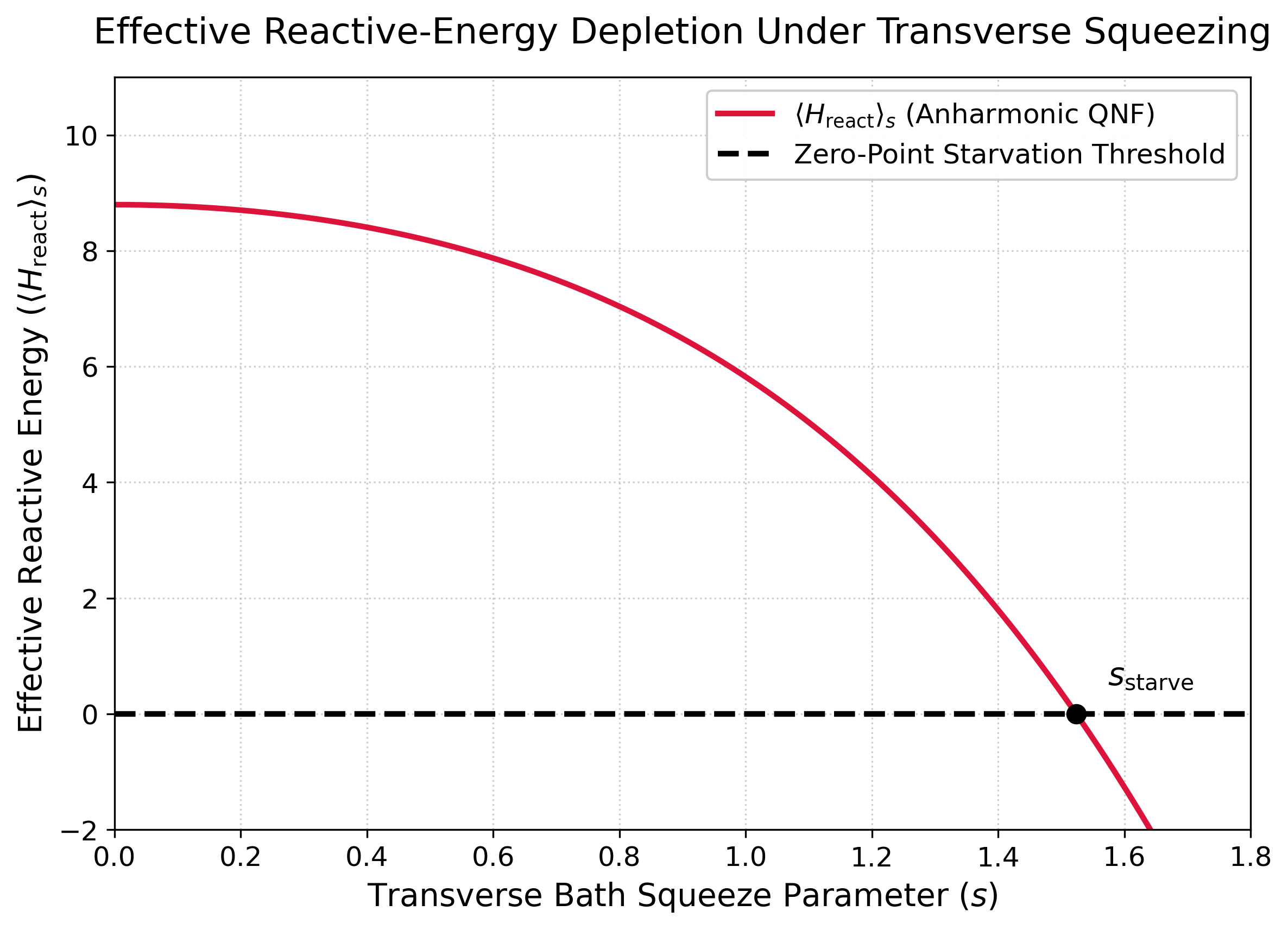}}{%
    \fbox{\parbox{0.68\textwidth}{\centering Missing file: quantum\_blockade\_fig.png}}}
    \caption{Effective reactive-energy diagnostic as a function of the
    transverse bath squeeze parameter $s$, computed by Wick--Isserlis
    evaluation of the truncated quantum normal form.  As $s$ increases,
    the expected bath actions grow and the effective expectation energy
    available to the reaction coordinate, $\langle H_{\mathrm{react}}\rangle_s$,
    is depleted.  The figure should be read as a QNF energy-depletion
    diagnostic, not as a proof of a quantum non-squeezing theorem.}
    \label{fig:blockade}
\end{figure}

As shown in Figure~\ref{fig:blockade}, evaluating
Eq.~\eqref{eq:H_react_exact} gives a concrete energy-depletion diagnostic.
Increasing $s$ forces exponential growth of the bath moments
$\langle J_2\rangle_s$ and $\langle J_2^2\rangle_s$.  At fixed total
expectation energy, this reduces $\langle H_{\mathrm{react}}\rangle_s$ and
can drive it through zero.  Once this happens, the corresponding Kemble
transmission diagnostic is exponentially small.

The local nature of the QNF must also be kept in view.  For large squeeze
parameters, the covariance distribution samples increasingly large
bath-action values.  The Wick--Isserlis moments remain exact for the
truncated polynomial, but the truncated QNF itself is only a local normal
form.  Therefore extreme squeezing may move the diagnostic outside the
physical domain where the local QNF is quantitatively reliable.

\section{Relative Squeeze Suppression Metric}

The principal transmission diagnostic is the relative squeeze suppression
metric.  It compares the transmission associated with a squeezed incoming
state to that of an isotropic minimum-uncertainty reference state at the
same total expectation energy.  This normalization separates the effect of
bath-mode squeezing from the baseline difficulty of tunneling through the
reactive barrier.

Let $T(E,s)$ denote a transmission diagnostic for an incoming Gaussian
state with total expectation energy $E$ and squeeze parameter $s$, with
$s=0$ corresponding to the isotropic reference state.  We define
\begin{equation}
    \mathcal S(E,s)=\frac{T(E,s)}{T(E,0)}.
    \label{eq:S_metric}
\end{equation}
A sharp decline of $\mathcal S(E,s)$ indicates that squeezing the bath mode
has suppressed transmission relative to the isotropic state at the same
energy.

For the quadratic model, $T(E,s)$ is the exact baseline transmission
$T_{\mathrm{quad}}(E,s)$ in Eq.~\eqref{eq:T_quad}.  For the anharmonic QNF
diagnostic, we map the effective reactive energy
$\langle H_{\mathrm{react}}\rangle_s$ through the one-dimensional Kemble
factor:
\begin{equation}
    T_{\mathrm{QNF}}(E,s)
    =\left[1+\exp\left(-\frac{2\pi}{\hbar\lambda}
      \langle H_{\mathrm{react}}\rangle_s\right)\right]^{-1}.
    \label{eq:T_QNF}
\end{equation}
The resulting ratio
\begin{equation}
        \mathcal S_{\mathrm{QNF}}(E,s)
        =\frac{T_{\mathrm{QNF}}(E,s)}{T_{\mathrm{QNF}}(E,0)}
\end{equation}
is an operational diagnostic of squeeze-induced suppression in the
truncated-QNF model.  It should be interpreted together with the action
scale $a_{\mathrm{sq}}(s)$ in Eq.~\eqref{eq:a_sq} and the classical
candidate width $c_{\mathrm{cand}}(E)$.

\begin{figure}[ht]
    \centering
    \IfFileExists{relative_suppression_fig.png}{%
    \includegraphics[width=0.7\textwidth]{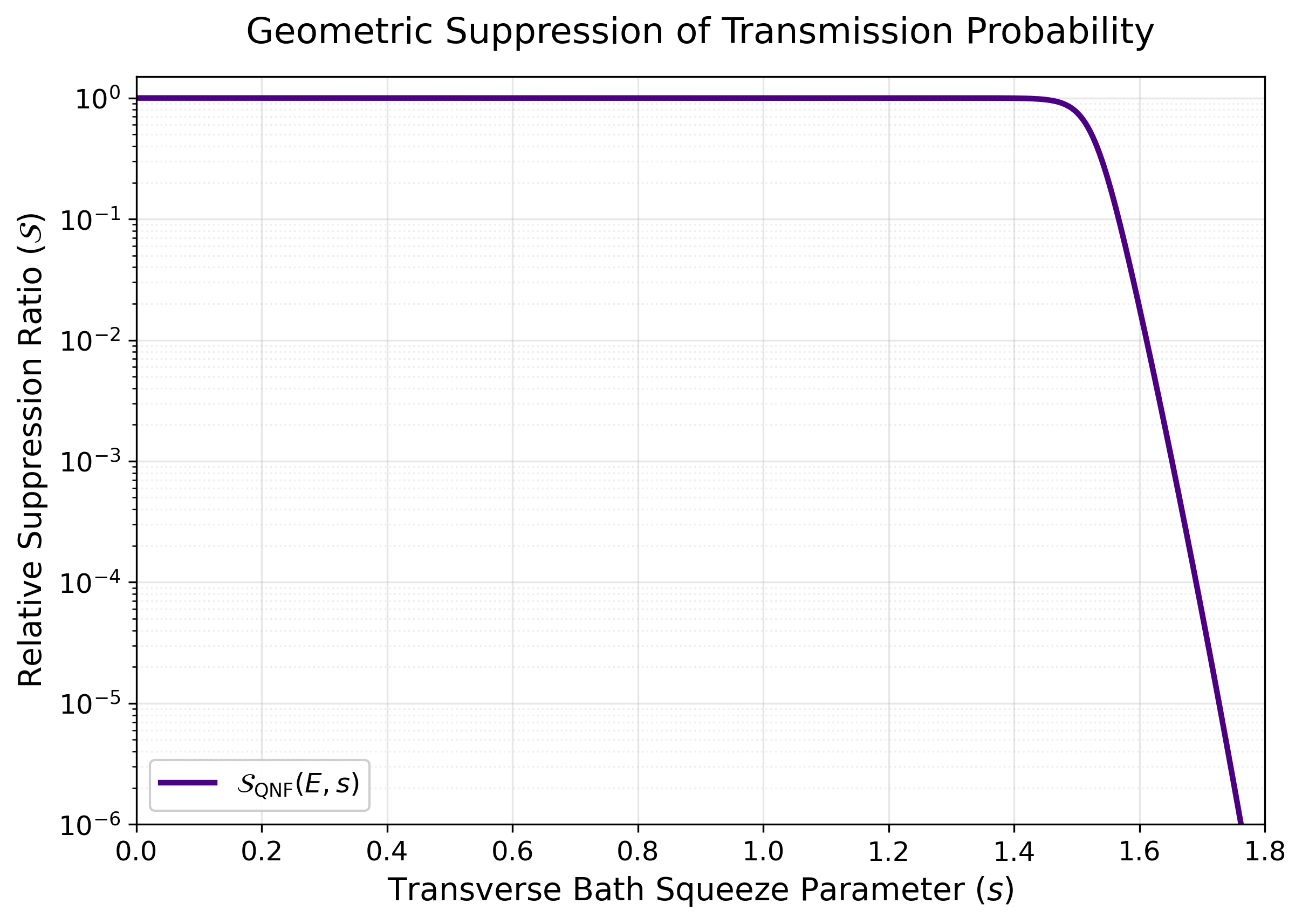}}{%
    \fbox{\parbox{0.68\textwidth}{\centering Missing file: relative\_suppression\_fig.png}}}
    \caption{Relative squeeze suppression metric
    $\mathcal S_{\mathrm{QNF}}(E,s)$ for the anharmonic QNF diagnostic.  The
    transmission diagnostic for the squeezed state is normalized by the
    isotropic state at the same total expectation energy.  The decay of
    the ratio quantifies squeeze-induced transmission suppression associated
    with growth of the bath-action expectation in the normal-form coordinates.}
    \label{fig:suppression}
\end{figure}

Figure~\ref{fig:suppression} illustrates this diagnostic.  The decay of
$\mathcal S_{\mathrm{QNF}}(E,s)$ is not presented as a universal quantum
reaction-rate theorem.  It shows that, in the QNF model, covariance-induced
growth of the bath-action expectation can strongly reduce the effective
reactive energy and therefore the associated transmission diagnostic.

\section{Physical Interpretation and Implications}

The calculations support a specific interpretation.  Quantum transmission
through a normal-form reaction bottleneck depends not only on the total
expectation energy of the incoming state, but also on how the state's
covariance geometry is distributed relative to the transverse bath actions.
A strongly squeezed state can have a large bath-action expectation value
even while remaining a minimum-uncertainty Gaussian state.  At fixed total
expectation energy, this increased bath-action contribution reduces the
effective energy available to the reactive degree of freedom.

\subsection{A geometric suppression mechanism}

The most precise description of the effect is a geometric suppression
mechanism.  The quantum transition state is not an impenetrable
symplectic wall, and the calculation does not establish an absolute
quantum filter.  Instead, when the squeezed-state bath-action scale becomes
large relative to the classical candidate bottleneck width, the normal-form
energy budget is increasingly allocated to the transverse mode.  The
reactive expectation energy is then reduced, and the corresponding
transmission diagnostic becomes exponentially small.

\subsection{Transverse modes as active participants}

Transverse bath modes are often treated as secondary degrees of freedom.
The present framework suggests a more active role.  Even if a bath mode is
linearly separable in the quadratic approximation, its covariance geometry
contributes to the action budget against which the normal-form bottleneck
is tested.  In this sense, transverse quantum preparation can influence
transmission through its effect on bath-action expectations.

\subsection{Suppressed transmission and residual tails}

Classically, exceeding a candidate width scale may be associated with
strong finite-time rejection or delay of an ensemble, depending on the
Hamiltonian dynamics.  Quantum mechanically, one should not expect a sharp
hard cutoff.  Even when the mean reactive energy is depleted, the incoming
state may have energy-distribution tails that contribute to transmission.
Thus the appropriate quantum language is suppression rather than absolute
blocking.  A complete scattering calculation would have to account for the
full energy distribution of the squeezed state, not only its mean reactive
energy.

\section{Conclusion}

We have developed a quantum normal-form diagnostic for squeeze-induced
transmission suppression near an index-1 saddle.  The construction is a
quantum companion to the classical symplectic-filter framework, but it is
not a proof of a quantum non-squeezing theorem for reactions.

In the quadratic saddle--center model, the transmission calculation is
exact within the separable model: the squeezed-state number distribution
in the bath mode is convolved with the one-dimensional Kemble transmission
factor for the reactive coordinate.  In the anharmonic truncated-QNF
setting, the central result is an exact Gaussian expectation-value
diagnostic for the truncated Weyl symbol.  Wick--Isserlis formulas give
closed algebraic expressions for the squeezed-state bath moments, and the
fixed total expectation energy then determines the effective reactive
energy.

The main physical conclusion is that squeezing a transverse bath mode can
produce strong transmission suppression by increasing the bath-action
expectation and depleting the reactive energy budget.  The relevant
quantum geometric scale is the covariance-based action-area diagnostic
$a_{\mathrm{sq}}(s)=2\pi\langle \hat J_2\rangle_s$, not a symplectic
capacity of the quantum state.  Comparing this diagnostic with the
classical candidate width scale $c_{\mathrm{cand}}(E)$ provides a precise
way to connect squeezed-state covariance geometry with the classical
normal-form bottleneck.

Several open problems remain.  First, the expectation-energy diagnostic
should be complemented by a treatment of the full energy distribution of a
highly squeezed state.  Second, the factorization
$\langle I J_2\rangle_s=\langle I\rangle_s\langle J_2\rangle_s$ is exact
for the incoming product state used here, but a full scattering theory
would have to track correlations generated by the interacting dynamics.
Third, a rigorous quantum analogue of the classical candidate-width
picture would likely require a covariance-ellipsoid or symplectic-eigenvalue
analysis, in the spirit of de Gosson's symplectic approach to quantum
mechanics \cite{deGosson2006,deGosson2009}.  These issues define the next
steps toward a more complete quantum geometric theory of reaction
bottlenecks.

\end{document}